\documentclass{ws-p8-50x6-00}
\usepackage{graphicx,wrapfig}
\begin{document}

\pagestyle{empty}

\begin{flushleft}
\Large{SAGA-HE-181-01    \hfill September 12, 2001}
\end{flushleft}
\vspace{1.5cm}
 
\begin{center}
\LARGE{\bf Determination of \\ 
           Parton Distribution Functions \\ in Nuclei} \\
\vspace{1.2cm}
\Large{ S. Kumano $^*$}  \\
\vspace{0.5cm}
{Department of Physics \\
 Saga University \\
 Saga, 840-8502, Japan} \\
\vspace{1.5cm}
\Large{Talk given at the 6th Workshop on} \\
{Non-Perturbative Quantum Chromodynamics} \\

\vspace{0.3cm}
{Paris, France, June 5 - 9, 2001} \\
{(talk on June 8, 2001) }  \\
\end{center}
\vspace{2.0cm}
\noindent{\rule{6.0cm}{0.1mm}} \\
\vspace{-0.3cm}
\normalsize

\noindent
{* Email: kumanos@cc.saga-u.ac.jp, \ URL: http://hs.phys.saga-u.ac.jp.}  \\

\vspace{+0.1cm}
\hfill {\large to be published in proceedings}
\vfill\eject
\setcounter{page}{1}
\pagestyle{plain}


\title{Determination of \\ Parton Distribution Functions in Nuclei}

\author{S. Kumano}

\address{Department of Physics, Saga University,
              Saga, 840-8502, Japan \\
              Email: kumanos@cc.saga-u.ac.jp,
              URL: http://hs.phys.saga-u.ac.jp}

\maketitle

\abstracts{Nuclear parton distribution functions are obtained
           by a $\chi^2$ analysis of lepton deep inelastic experimental data.
           It is possible to determine valence-quark distributions
           at medium $x$ and antiquark distributions at small $x$;
           however, the distributions in other $x$ regions and gluon
           distributions cannot be fixed.
           We need a variety of experimental data
           and also further analysis refinements.}

\section{Introduction}\label{intro}

Nuclear parton distributions are used inevitably for calculating high-energy
nuclear cross sections; however, precise distributions are not obtained yet.
On the other hand, heavy-ion reactions have been investigated for
finding a quark-gluon plasma signature. Because it should be found
in any unusual cross section which cannot be explained by 
the present hadron physics framework, the parton distributions have to
be known very precisely. Although there are many studies on quark-gluon
plasma signatures, it is unfortunate that the same amount of efforts 
are not made for the initial condition, namely the parton distributions.
In fact, many researchers just use the parton distributions
in the ``nucleon" instead of those in nuclei.
It is known that nuclear distributions are modified from those in the nucleon,
and the modification could be of the order of 20\%.
However, little information is available for nuclear gluon
distributions, which play a crucial role for the $J/\psi$ production. 

The nuclear parton distributions were investigated, for example,
in Ref.\,1, and the first $\chi^2$ analysis was reported in
Ref.\,2. However, it should be noted that the $\chi^2$ analysis is
still at the preliminary stage in comparison with many solid investigations
on the distributions in the nucleon.
There are two major issues. First, there are not many available
data for nuclei. In particular, the data come mainly from deep inelastic
electron or muon scattering. Second, the technique of nuclear $\chi^2$
analysis is not established. The studies in Ref.\,2 tried to set up a $\chi^2$
analysis method for the nuclear distributions.

In this paper, the optimum distributions obtained in Ref.\,2 are explained.
The analysis method is discussed in Sec.\,\ref{method}.
Results are shown in Sec.\,\ref{results}, and a parton distribution library
is explained in Sec.\,\ref{codes}.
Our studies are summarized in Sec.\,\ref{summary}.

\section{Analysis method}\label{method}

The nuclear parton distributions are defined at a fixed $Q^2$, which is 
taken 1 GeV$^2$ ($\equiv Q_0^2$). This $Q^2$ point is selected so that
many experimental data become available and yet perturbative QCD
is expected to be applied. 
Then, the nuclear distributions are defined by $x$ dependent functions
multiplied by the distributions in the nucleon.
It is analogous to polarized distributions. In the leading order (LO),
the polarized distributions are restricted by the positivity condition.
Namely, they should be smaller than the unpolarized ones.
The positivity condition is easily handled if the polarized distributions
are defined from the unpolarized.\cite{aac}
In a similar way,  it is technically easier to parametrize
the modification part instead of the nuclear distributions themselves,
because the nuclear modification is typically smaller than 20\%.
The nuclear parton distributions are then given as
\begin{equation}
f_i^A (x, Q_0^2)  = w_i(x,A,Z) \, f_i (x, Q_0^2)  ,
\end{equation}
where $f_i (x, Q_0^2)$ is the $i$-type distribution in the nucleon,
and $w_i(x,A,Z)$ is a weight function. As the distribution types,
$i$=$u_v$, $d_v$, $\bar q$, and $g$ are taken.
The distributions in the nucleon are taken from
the MRST-LO parametrization.\cite{mrst98}
The nuclear modification part $w_i(x,A,Z)$ is parametrized as
\begin{equation}
w_i(x,A,Z)  = 1+\left( 1 - \frac{1}{A^{1/3}} \right) 
          \frac{a_i(A,Z) +b_i x+c_i x^2 +d_i x^3}{(1-x)^{\beta_i}}  ,
\label{eqn:wi}
\end{equation}
in terms of the parameters $a_i$, $b_i$, $c_i$, $d_i$, and $\beta_i$.
At this stage, a simple $A$ dependent form ($\sim A^{1/3}$) is
assumed\,\cite{sd} in order to avoid complexity.
The function $1/(1-x)^{\beta_i}$ is introduced
so as to reproduce the Fermi motion part at large $x$. 
The rest of the $x$ dependence is assumed in the cubic functional form,
so that this $\chi^2$ analysis is called a ``cubic" type.
We also tried another simpler one, a ``quadratic" type, without
the $d_i x^3$ term in Eq. (\ref{eqn:wi}).
There are three constraints for the parameters:
the conditions for nuclear charge, baryon number, and momentum.
Therefore, three parameters can be fixed.

Experimental data are taken from those in electron or muon deep inelastic
scattering (DIS). The analysis with other data is in progress.
There are also neutrino-nucleus data. However, it is difficult
to address ourselves to the nuclear modification because there is no accurate
data for the neutrino-deuteron scattering. The situation will change
if a neutrino factory is materialized.\cite{sknu}

The data are taken at various $Q^2$ points. The initial 
parton distributions are evolved to the experimental $Q^2$ points so as to
calculate $\chi^2$:
\begin{equation}
\chi^2 = \sum_j \frac{(R_{F_2,j}^{A,data}-R_{F_2,j}^{A,theo})^2}
                     {(\sigma_j^{data})^2},
\label{eqn:chi2}
\end{equation}
where $R_{F_2} ^A (x,Q^2) = F_2^A (x,Q^2) / F_2^D (x,Q^2)$.
The analysis is done in the leading order of $\alpha_s$, so that
the structure function $F_2^A$ is given by
\begin{equation}
F_2^A (x,Q^2) = \sum_q e_q^2 x [ q^A(x,Q^2) + \bar q^A(x,Q^2) ],
\end{equation}
where $e_q$ is the quark charge,
and $q^A$ ($\bar q^A$) is the quark (antiquark) distribution
in the nucleus $A$.
The data exist for various nuclei, which are assumed as
$^4$He, $^7$Li, $^9$Be, $^{12}$C, $^{14}$N, $^{27}$Al, $^{40}$Ca,
$^{56}$Fe, $^{63}$Cu, $^{107}$Ag, $^{118}$Sn,
$^{131}$Xe, $^{197}$Au, and $^{208}$Pb
in the theoretical analysis. The $Q^2$ evolution is calculated by
the ordinary leading-order DGLAP equations. 

\section{Results}\label{results}

\begin{table}[b]
\caption{$\chi^2$ contributions.}
\label{tab:chi2}
\begin{center}
\footnotesize
\begin{tabular}{|c|c|c|c|}
\hline
nucleus & \# of data & $\chi^2$ (quad.)  & $\chi^2$ (cubic)  \\
\hline\hline
He    &  35  &  \ 55.6   &  \     54.5  \\
Li    &  17  &  \ 45.6   &  \     49.2  \\ 
Be    &  17  &  \ 39.7   &  \     38.4  \\
C     &  43  &  \ 97.8   &  \     88.2  \\
N     & \ 9  &  \ 10.5   &  \     10.4  \\
Al    &  35  &  \ 38.8   &  \     41.4  \\
Ca    &  33  &  \ 72.3   &  \     69.7  \\
Fe    &  57  &   115.7   &  \     92.7  \\
Cu    &  19  &  \ 13.7   &  \     13.6  \\
Ag    & \ 7  &  \ 12.7   &  \     11.5  \\
Sn    & \ 8  &  \ 14.8   &  \     17.7  \\
Xe    & \ 5  & \ \ 3.2   &  \  \ \ 2.4  \\
Au    &  19  &  \ 55.5   &  \     49.2  \\
Pb    & \ 5  & \ \ 7.9   &  \  \ \ 7.6  \\	
\hline
total & 309  &   583.7   &  \    546.6  \\
\hline
\end{tabular}
\end{center}
\end{table}
\vspace{0.0cm}
\normalsize

The $\chi^2$ analysis was done with the help of the CERN {\sc Minuit}
subroutine. The detailed descriptions of the analysis should be
found in Ref.\,2. The obtained $\chi^2$ values are listed in
Table \ref{tab:chi2}. The table indicates that the fit is not excellent
in lithium, carbon, calcium, iron, and gold. In comparison with
the quadratic analysis, the fit becomes better
notably for carbon, iron, and gold in the cubic type; however,
by sacrificing $\chi^2$ values of some other nuclei.
The number of the data is 309, so that
$\chi^2$ per degrees of freedom is given by
$\chi_{min}^2/d.o.f.$=1.93 (quadratic) or 1.82 (cubic).
Because they are certainly much larger than one, they may not seem to be
excellent fits. However, it is mainly due to the scattered experimental
data.

\vspace{-0.0cm}
\noindent
\begin{figure}[h]
\parbox[t]{0.48\textwidth}{
   \begin{center}
\includegraphics[width=0.48\textwidth]{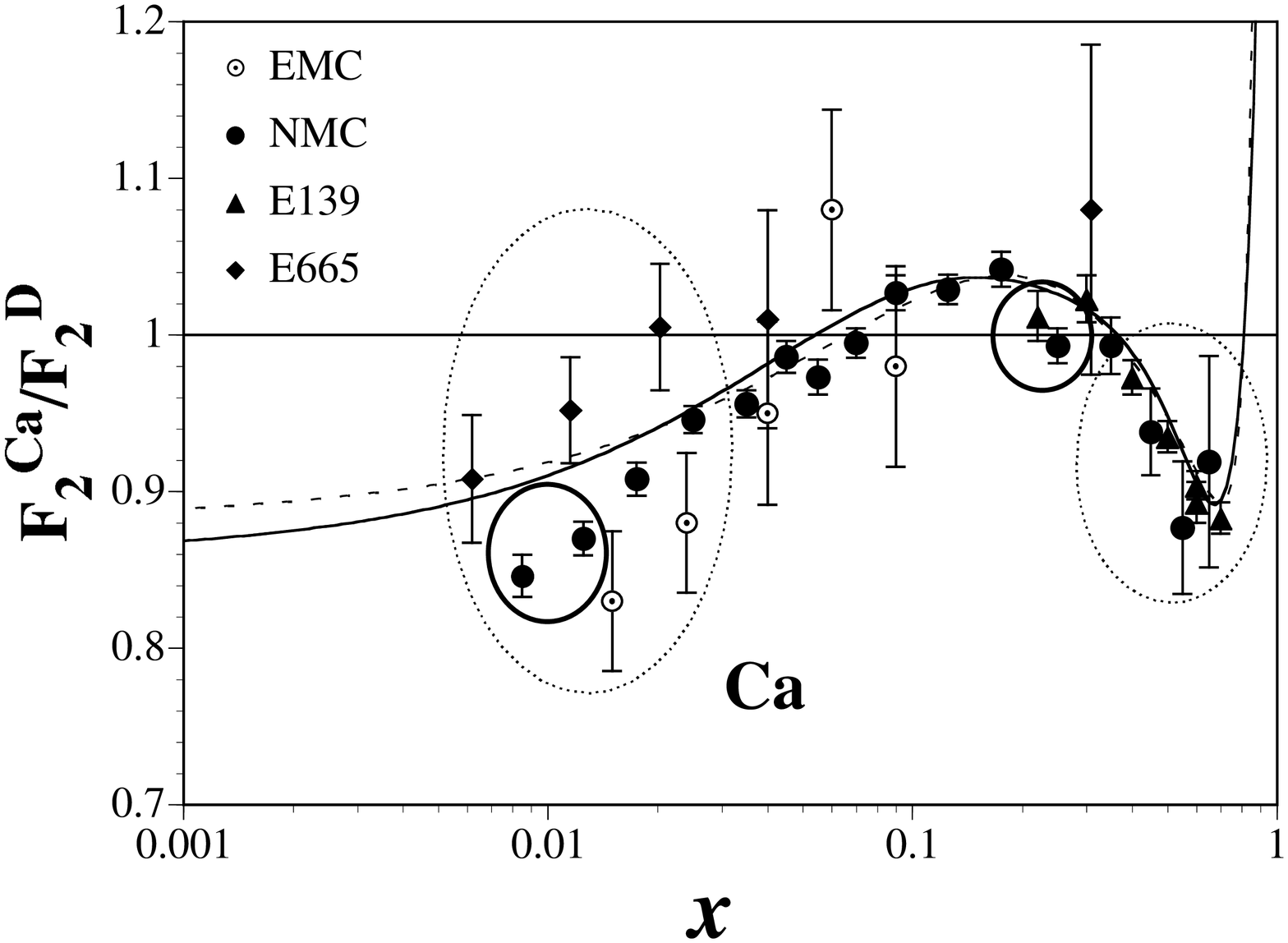}
   \end{center}
   \vspace{-0.4cm}
       \caption{\footnotesize Comparison with calcium data.}
       \label{fig:ca-chi2}
}\hfill
\parbox[t]{0.48\textwidth}{
   \begin{center}
\includegraphics[width=0.48\textwidth]{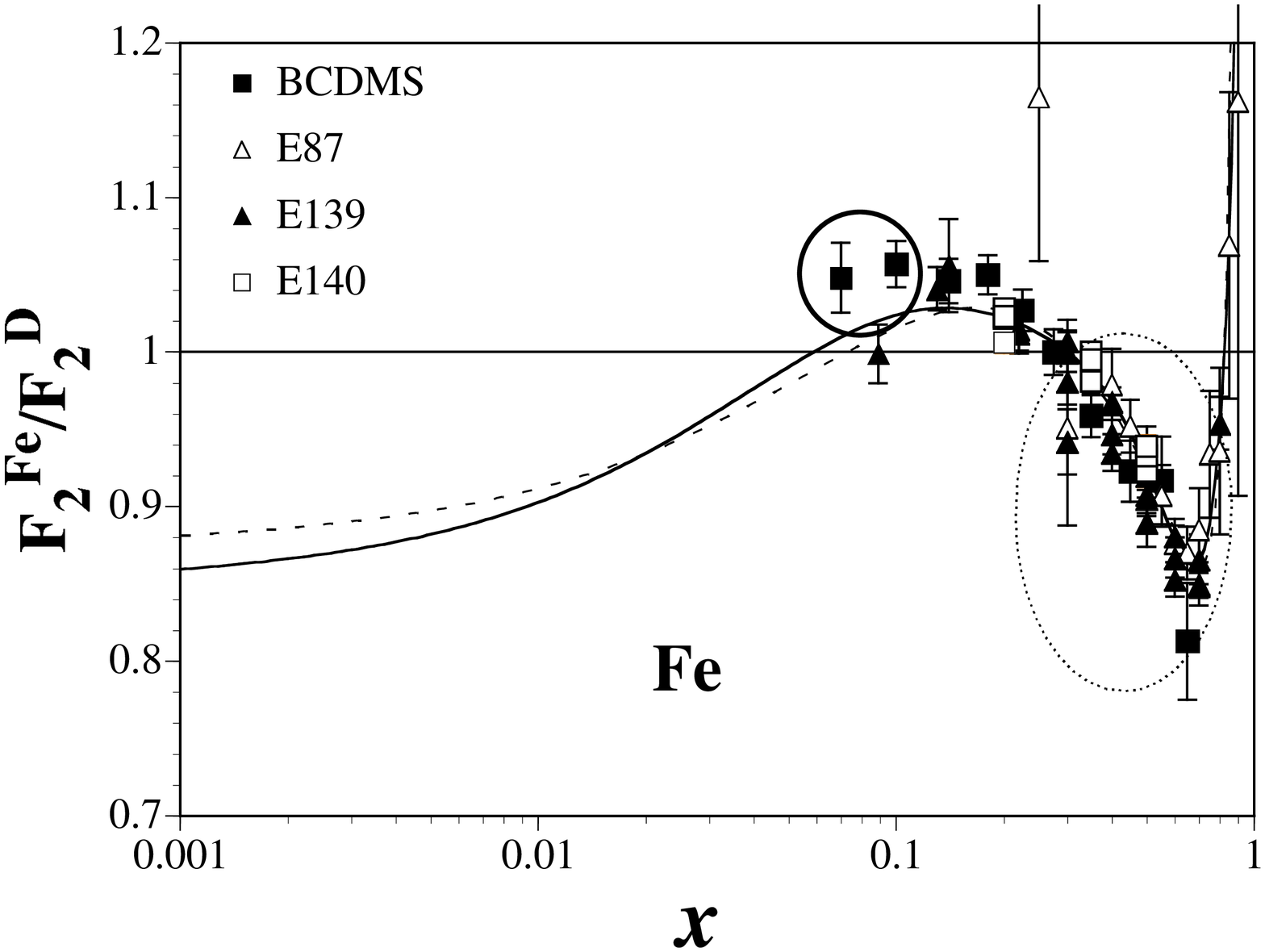}
   \end{center}
   \vspace{-0.4cm}
       \caption{\footnotesize Comparison with iron data.}
       \label{fig:fe-chi2}
}
\end{figure}
\vspace{-0.0cm}

In order to specify the origin of the large $\chi^2$ contributions,
we show our fitting results in comparison with the data in
Figs. \ref{fig:ca-chi2} and \ref{fig:fe-chi2}, where the solid
circles indicate the major source of the $\chi^2$ contributions and
the dotted ones indicate other sources.
Our fitting results are shown by the dashed and solid curves 
for the quadratic and cubic analyses, respectively, and they
are calculated at $Q^2$=5 GeV$^2$. Although they cannot be directly
compared with the data due to the $Q^2$ difference,
the figures suggest that the fits should be well done. 
Both results are almost the same except for the small $x$ region, where
the data do not exist.
It is obvious from these figures that the data are scattered
and some of them have very small errors, which contribute mostly
to the total $\chi^2$.
Even if an excellent $\chi^2$ analysis method is developed in future
by introducing complicated $A$ and $x$ dependence, it is inevitable
to obtain $\chi^2 /d.o.f. > 1$ in the present experimental situation.

\begin{wrapfigure}{r}{0.46\textwidth}
   \vspace{+0.5cm}
     \begin{center}
     \includegraphics[width=0.45\textwidth]{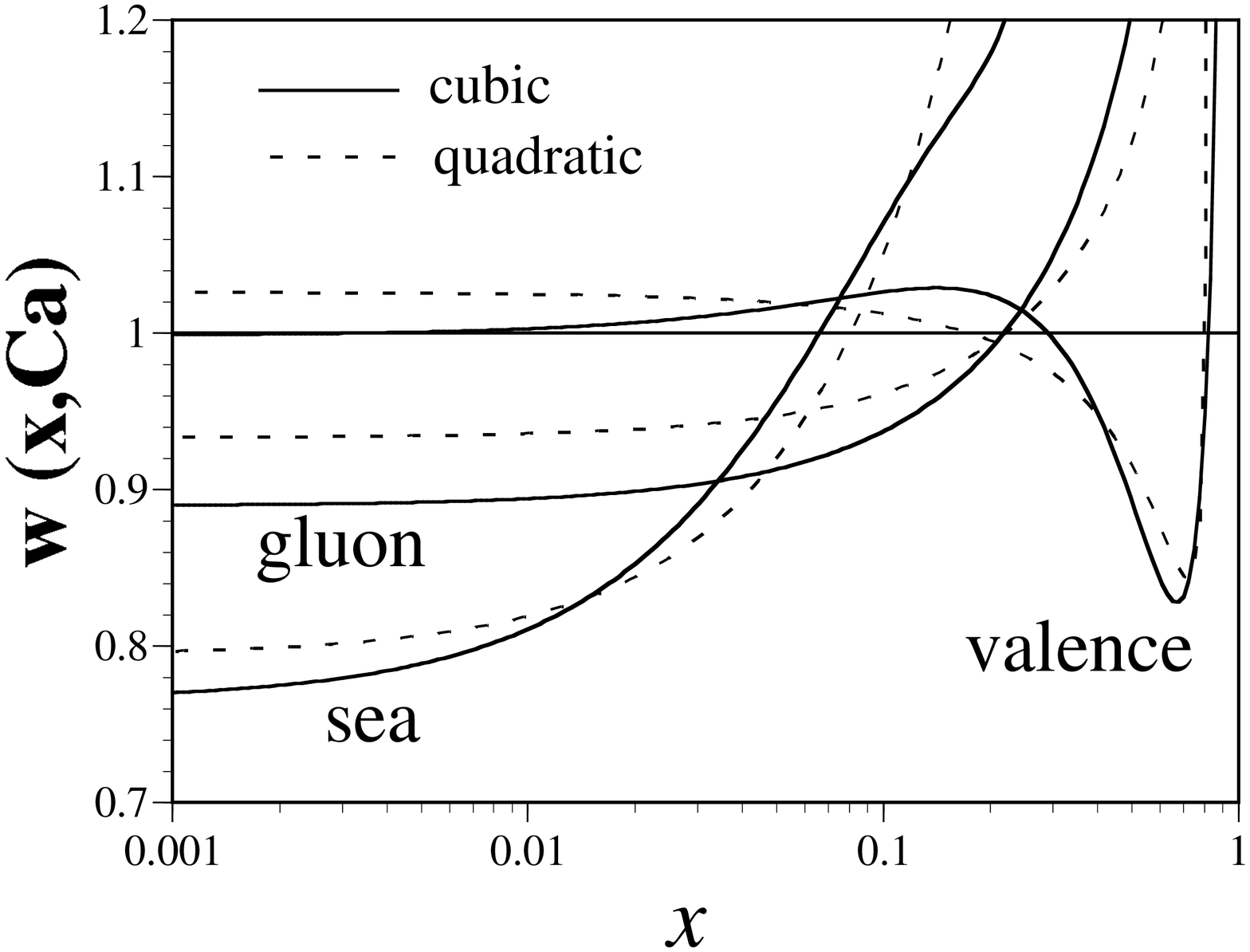}
     \end{center}
 \vspace{-0.4cm}
   \caption{\footnotesize Weight functions.}
   \label{fig:wax}
\end{wrapfigure}
Next, obtained weight functions are shown for the calcium nucleus
in Fig. \ref{fig:wax}, where the dashed and solid curves indicate
the results for the quadratic and cubic analysis, respectively,
at $Q^2$=1 GeV$^2$. As expected, the valence-quark distributions
explain the EMC effect at medium $x$ and they have Fermi-motion-type
increase at large $x$. However, the small $x$ behavior is far from
obvious. It could show either antishadowing or shadowing.
A precise determination of the valence distributions, especially
at small $x$, should be done by a future neutrino factory.\cite{sknu}

The antiquark distributions at small $x$ are restricted by
the $F_2^A$ shadowing, so that they could be fixed in this $x$ region.
However, they cannot be determined at medium and large $x$.
If the Drell-Yan data are added to the analysis, we expect to have
more restriction on the antiquark distributions at $x \sim 0.1$.

On the other hand, the gluon distribution cannot be determined well
in the present analysis. The inclusive DIS data are not sensitive
to the gluon distributions, especially in the LO analysis.
At this stage, the gluon distributions seem to show shadowing
at small $x$, and they increase at large $x$ because of the momentum
conservation. In future, we should consider to include the data which could
restrict the gluon distributions.

\section{Parton distribution codes}\label{codes}

From the $\chi^2$ analysis, the parton distributions are obtained for
nuclei from the deuteron to a large nucleus with $A \sim 208$.
Because variations are small from  $A=208$ to nuclear matter,
we expect the distributions could be extrapolated into larger $A$ $(> 208)$.
We set up the initial distributions at $Q^2$=1 GeV$^2$. However,
we made it possible to calculate the distributions at any $Q^2$
by our computer codes, because the $Q^2$ evolution may be tedious
for some users. The codes can be obtained from our web site.\cite{nucl-lib}
There are two possibilities of using our results for the parton distributions
in a user's project. One is to use the analytical expressions for the weight
functions, and another is to use the computer codes.
Strictly speaking, the obtained distributions are valid for
the analyzed nuclei, helium, lithium, $\cdot \cdot \cdot$, and lead.
However, the $A$ dependence is reproduced well even by the simple
$1/A^{1/3}$ form, so that our studies are expected to be used also
for other nuclei except for unstable ones with large neutron excess.

\vspace{0.3cm}
\noindent
\underline{Analytical expressions}
\vspace{0.15cm}

For those who have own $Q^2$ evolution codes, we wrote the analytical
expressions of the weight functions in Appendix of Ref.\,2.
They should be multiplied by the MRST-LO distributions in the nucleon
so as to obtain the nuclear distributions at $Q^2$=1 GeV$^2$.
Then, they should be evolved to a $Q^2$ point in a user's project.
Because the parton distributions in the nucleon are similar
among various groups, the results are not expected to change
significantly even if other parametrization is used instead
of the MRST-LO.

\vspace{0.3cm}
\noindent
\underline{Computer codes}
\vspace{0.15cm}

For those who are not familiar with the $Q^2$ evolution,
we supply our own codes for calculating the nuclear parton distributions
at any $Q^2$. Grid data are prepared for the distributions with
the variables $x$ and $Q^2$, and they are interpolated.
If one wishes to run the actual evolution, we also supply a $Q^2$
evolution package. The detailed instructions
should be found in the distributed file, saga01.tar.gz.\cite{nucl-lib}

\section{Summary}\label{summary}

We have obtained optimum parton distributions in nuclei by the $\chi^2$
analysis of DIS experimental data. At this stage, our studies are intended
to set up a tool for the nuclear $\chi^2$ analysis, which had not been
done at all until recently. It is still far from completion in the sense
that analysis refinements are needed and a variety of experimental data
should be included in the analysis. We continue to work on this project
for obtaining reliable parton distributions in nuclei.

\section*{Acknowledgments}
S.K. was supported by the Grant-in-Aid for Scientific
Research from the Japanese Ministry of Education, Culture, Sports,
Science, and Technology. 
This talk is based on the results in Ref.\,2, where the parametrization was
investigated with M. Hirai and M. Miyama.



\end{document}